\documentclass[fleqn,usenatbib]{mnras}

\usepackage{newtxtext,newtxmath}
\usepackage{enumitem}
\usepackage[normalem]{ulem}
\usepackage{subcaption} 
\usepackage{placeins}
\usepackage[T1]{fontenc} 

\DeclareRobustCommand{\VAN}[3]{#2}
\let\VANthebibliography\thebibliography
\def\thebibliography{\DeclareRobustCommand{\VAN}[3]{##3}\VANthebibliography}

\usepackage{graphicx}
\usepackage{subcaption}	% Including figure files
\usepackage{amsmath}	% Advanced maths commands
\usepackage{algorithm}
\usepackage{algpseudocode}
\usepackage{xcolor}
\usepackage{float}
\usepackage{booktabs}
\usepackage{tikz}
\usetikzlibrary{positioning, arrows.meta, shapes.geometric, calc, fit}

\newcommand*    \msun{{\,\rm{M}_{\odot}}}

\newcommand*    \kpc{{\,\mathrm{kpc}}}

\newcommand*    \mbh{M_{\rm bh}}

\newcommand*    \mbulge{M_{\rm b}}

\title[NODEs for BHBs]
{Continuous-Time Modelling of Black Hole Binary Evolution with Neural ODEs}

\author[Julian Chan et al.]{
Julian Chan$^{1}$\thanks{E-mail: julianjunyen.chan@surrey.ac.uk},
Alessia Gualandris$^{1}$,
Payel Das$^{1}$
\\
$^{1}$School of Mathematics and Physics, Faculty of Engineering and Physical Sciences, 
University of Surrey, Guildford GU2 7XH, UK\\
}

\date{Accepted XXX. Received YYY; in original form ZZZ}

\pubyear{2024}

\begin{document}
\label{firstpage}
\pagerange{\pageref{firstpage}--\pageref{lastpage}}
\maketitle

\begin{abstract}
Pulsar timing arrays (PTAs) can detect the low-frequency stochastic gravitational-wave background (GWB) generated by an ensemble of supermassive black hole binaries (BHBs). Accurate determination of BHB merger timescales is essential for interpreting GWBs and constraining key astrophysical quantities such as black hole (BH) occupation fractions and galaxy coalescence rates. High-accuracy $N$-body codes such as \texttt{Griffin} can resolve sub-pc BHB dynamics but are too costly to explore a wide range of initial conditions, motivating the need for surrogate models that emulate their long-term evolution at much lower computational cost. We investigate neural ordinary differential equations (NODEs) as surrogates for the secular orbital evolution of BHBs. Our primary contribution is a parameterised NODE (PNODE) trained on an ensemble of $N$-body simulations of galaxy mergers spanning a two-dimensional parameter space defined by the initial orbital eccentricity and particle resolution $(e_i, N)$, with the learned vector field explicitly conditioned on these parameters. A single PNODE thereby learns a simulation-parameter-conditioned dynamical model for the coupled evolution of the BH pair's orbital state across the ensemble, yielding smooth trajectories from which stable hardening and eccentricity growth rates can be extracted. The PNODE accurately reproduces the secular evolution of the specific orbital energy and angular momentum, and the corresponding Keplerian orbital elements, for held-out trajectories, with modest generalisation to a partially unseen high-resolution case. Combining PNODE predictions with semi-analytical prescriptions for stellar hardening and gravitational-wave emission yields BHB merger timescales consistent with those obtained from direct $N$-body inputs within current theoretical uncertainties.
\end{abstract}


%%%%%%%%%%%%%%%%%%%%%%%%%%%%%%%%%%%%%%%%%%%%%%%%%%

%%%%%%%%%%%%%%%%% BODY OF PAPER %%%%%%%%%%%%%%%%%%

\section{Introduction}
Supermassive black holes (SMBHs) are ubiquitous in the centres of massive galaxies~\citep{ferrarese_supermassive_2005,kormendy_coevolution_2013}, and their masses obey scaling relations with properties of their hosts, such as the bulge mass and the stellar velocity dispersion~\citep{gebhart_2000,ferrarese_meritt_2000, kormendy_coevolution_2013,haring_rix_2004,mcconell_2013}. These scaling relations are widely interpreted as evidence for a co-evolutionary history between SMBHs and their host galaxies, suggesting that BHs and galactic bulges are ultimately linked~\citep{heckman2014coevolution, kormendy_coevolution_2013}. Within the $\Lambda$CDM cosmological model for structure formation, galaxies grow
hierarchically via a ``bottom-up'' process through repeated mergers, progressively forming larger structures~\citep{white_rees_1978}. 
As a result of such mergers, the SMBHs are driven towards the centre of the merger remnant, where they eventually form a supermassive black hole binary (BHB)~\citep{begelman_1980}. The journey of SMBHs from the galactic merger to binary formation spans multiple time and length scales. Typically, it consists of three distinct phases: 1) the dynamical friction phase, in which interactions with the stellar and dark matter components of the galaxy induce a drag force on the SMBHs, causing them to sink towards the centre of the merger remnant and form a bound pair~\citep{chandrasekhar_1943}; 2) the binary hardening phase, in which the bound pair experiences a series of three-body interactions
with stars on low angular momentum orbits, also known as scatterings, which extract energy and angular momentum from the binary and transfer them to the stars, thereby leading to the excavation of a core in the stellar density profile; 3) the gravitational wave (GW) phase, during which the SMBHs inspiral due to emission of GWs until they reach coalescence to form a single SMBH. Binaries have been shown to efficiently reach the GW phase and coalesce within a Hubble time, provided that a sufficient supply of stars is available to interact with the BHBs~\citep{vasiliev_2015, gualandris_2017}. This is commonly achieved in remnants of mergers due to the triaxiality of the gravitational potential and the resulting diffusion of angular momentum that ensures sustained refilling of the loss cone~\citep{bortolas2018}. However, the time to coalescence varies widely, depending on the properties of galaxies and those of the BHBs at formation~\citep{khan_2018, mannerkoski2019, gualandris2022, fastidio2024}. 

BHBs are expected to be the loudest sources of GWs in the Universe and prime astrophysical candidates for the gravitational wave background (GWB) signal recently observed by Pulsar Timing Arrays (PTAs) in the nHz frequency regime~\citep{agazie_2023,antoniadis_2023,reardon_2023,xu_2023}. The signal emerges because of the incoherent superposition of the emission from a cosmic population of unresolved BHBs. Alternative origins for the GWB include inflation, phase transitions, cosmic strings, tensor mode generation by the nonlinear evolution of scalar perturbations in the early
Universe, and oscillations of the galactic potential in the presence of ultralight dark matter~\citep{EPTA2024}.

The astrophysical exploitation of PTA observations has the potential to constrain key quantities such as the coalescence rate of galaxies, the occupation fraction of SMBHs, the efficiency of BHB mergers and SMBH-galaxy scaling relations~\citep{antoniadis_2023}. Accurate predictions of merger timescales from numerical simulations of BHBs are crucial to breaking degeneracies and interpreting the observed GWB. Current estimates, mainly from large cosmological simulations or semi-analytical models, vary from a few $\mathrm{Myr}$ to several $\mathrm{Gyr}$ and possibly longer, depending on model parameters and assumptions~\citep[e.g.][]{Ricarte2019, Barausse2020, 2022IzquierdoVillalba, dong2023}. Evolutionary timescales are particularly sensitive to the eccentricity of BHBs at the time of binary formation, as both dynamical hardening through encounters with stars and GW inspiral depend on eccentricity~\citep{nasim2020}. It is unclear whether this is due to a higher sensitivity to resolution~\citep{nasim2020, gualandris2022, fastidio2024}, or whether it is intrinsic to binary pairing~\citep{rawlings2023}. Regardless of its origin, constraining the eccentricity of binaries at formation or even at hard-binary separation is key to interpreting the GWB signal. \citet{sesana_2013} demonstrated that eccentric binaries produce a flattening of the GWB spectrum at low frequencies, a feature that is potentially observable by PTAs~\citep{antoniadis_2023, EPTA2024}.

% Modelling galactic mergers from kpc to sub-pc scales with both high accuracy and high resolution, e.g. with $N$-body codes like \texttt{Griffin} ~\citep{dehnen_2014} and \textsc{KETJU}~\citep{rantala_2017, mannerkoski2023}, is computationally challenging. The large computational complexity limits the resolution that can be realistically employed, and frequently leads to the choice of short integration times, often barely reaching the end of the fast hardening phase. Limited resolution results in stochasticity of the binary eccentricity and noisy orbital parameter determination for the BHBs. 
% Furthermore, the orbital elements of the binary at the end of the $N$-body integration are generally used as input for semi-analytical models that integrate the motion of BHBs to coalescence. This allows us to estimate total evolutionary times from formation to coalescence, which in turn contribute to the calculation of the GWB amplitude detected by PTA and merger rates for other detectors, such as the Laser Interferometer Space Antenna (LISA). The semi-analytical models are solvers of coupled differential equations, and as such are sensitive to the initial conditions, i.e. the orbital elements of the BHB~\citep[e.g.][]{gualandris_2012, attard2024, fastidio2024}. 

Directly modelling galaxy mergers from kpc to sub-pc scales at both high accuracy and high resolution, for example with $N$-body codes such as \texttt{Griffin}~\citep{dehnen_2014} and \textsc{KETJU}~\citep{rantala_2017,mannerkoski2023}, remains computationally challenging. The high cost limits the resolution and integration time that can be achieved in practice and often leads to simulations that barely reach the end of the rapid hardening phase. Finite resolution and limited sampling lead to stochasticity in the binary eccentricity and noisy determinations of orbital parameters. Moreover, the orbital elements of the binary at the end of the $N$-body integration are typically used as initial conditions for semi-analytical models that evolve the BHB to coalescence via coupled differential equations~\citep[e.g.][]{gualandris_2012,attard2024,fastidio2024}. These semi-analytical models are sensitive to the initial orbital elements, so noise and systematic biases in the $N$-body outputs propagate into BHB merger timescales.

To address these challenges, we develop a neural ordinary differential equation (NODE) framework as a surrogate model for the secular evolution of BHBs inferred from $N$-body simulations. NODEs provide a continuous-time parametrisation of the underlying vector field that governs the evolution of a dynamical system, with an ODE-based inductive bias that favours smooth trajectories. As an initial test, we apply a NODE to the orbital evolution of a single massive BHB from a high-resolution $N$-body simulation and find that several baseline models perform comparably or even better in this scenario. We therefore focus on the more realistic setting of an ensemble of $N$-body simulations of galaxy mergers that sample a parameter space of initial galactic orbital eccentricity and particle resolution $(e_i, N)$, and we introduce a parametrised NODE (PNODE) conditioned on such parameters that learns a unified vector field whose dynamics vary smoothly across this region of parameter space. The PNODE successfully recovers the secular evolution of the specific orbital energy and angular momentum, together with the associated Keplerian orbital elements of the BHB, demonstrates modest extrapolation in resolution. Finally, by coupling PNODE-predicted orbital elements to a semi-analytical model thats describes stellar hardening and gravitational-wave emission, we find that the resulting merger timescales remain consistent with those obtained from direct $N$-body inputs within current theoretical uncertainties, suggesting that PNODEs can act as a simulation-parameter-aware surrogates.

\section{Machine learning Concepts}\label{sec:theory}
This section introduces the key machine learning concepts underpinning our methodology, focusing on the transition from discrete-depth neural architectures to continuous-depth formulations. We begin with a brief overview of canonical feedforward neural networks (FFNNs), and then describe neural ordinary differential equations (NODEs), which recast deep learning as a continuous-time dynamical process suited to modelling physical systems.

% Finally, we describe symbolic regression, an approach for extracting interpretable, closed-form relationships from learnt representations.

\subsection{Deep Learning} \label{sec:deep learning}
Deep learning is a subset of machine learning that uses deep neural networks with many layers of nonlinear transformations to learn hierarchical representations from data. By passing inputs through successive nonlinear layers, these models extract increasingly abstract features. Although deep architectures are especially effective for complex, high-dimensional problems, the underlying principles of neural representation learning, such as parameter sharing, nonlinearity and differentiability, are broadly applicable.

In the context of this study, we are only concerned with the supervised learning setting. The objective is to learn a function $f: \mathcal{X} \to \mathcal{Y}$ that maps inputs $\mathbf{x} \in \mathcal{X}$ to outputs $\mathbf{y} \in \mathcal{Y}$. Typically, the inputs $\mathbf{x}$ form a $D$-dimensional
vector where $\mathbf{x} \in \mathbb{R}^D$, are commonly known as features. For regression tasks, we are interested in predicting real vector-valued outputs $\mathbf{y} \in \mathbb{R}^{d}$, also known as targets. The function $f$ can be learnt from a training dataset $\mathcal{D} = \{ (\mathbf{x}_n, \mathbf{y}_n)\}^{N}_{n=1}$, using a variety of supervised learning algorithms. These may include linear models such as linear regression, polynomial regression for capturing nonlinear relationships, and logistic regression for classification tasks. Other popular methods include tree-based approaches, support vector machines, and neural networks. The selection of an optimal learning algorithm is problem-dependent, as the \textit{no-free-lunch} theorem posits that no single algorithm consistently outperforms all others across all supervised learning tasks. Thus, choosing the right algorithm requires careful consideration of the specific problem characteristics. 

\subsection{Discrete-Depth Architectures}\label{sec:discrete_depth}
In the modern deep learning paradigm, a neural network (NN) signifies any differentiable function expressible as a computation graph, where nodes correspond to fundamental operations, such as matrix multiplication, while edges represent the flow of numerical data in the form of vectors, matrices, or higher-order tensors. Most common NNs are discrete-depth architectures, characterised by a finite sequence of parameterised, differentiable transformations.

\subsubsection{Feedforward Neural Networks}\
A canonical example of a discrete-depth architecture is the feedforward neural network (FFNN), which consists of a composition of $L$ nonlinear operations $\mathbf{g}_{l}$, where $l = 1, 2, \ldots, L$, on intermediate representations $\mathbf{z}_{l}$. This can be represented succinctly by:
\begin{equation}\label{eq:ff_network}
	\mathbf{f}_{\boldsymbol{\theta}}(\mathbf{x}) = \mathbf{g}^{(L)}_{\boldsymbol{\theta}_L} \circ \mathbf{g}^{(L-1)}_{\boldsymbol{\theta}_{L-1}} \circ \dots \circ \mathbf{g}^{(1)}_{\boldsymbol{\theta}_1}(\mathbf{x}).
\end{equation}
Each $\mathbf{g}^{(l)}$ is specified by applying an affine transformation followed by a nonlinear activation function $\sigma$:
\begin{equation}
	\mathbf{g}_{\boldsymbol{\theta}_i}^{(l)}(\mathbf{z}^{(l)}) = \sigma_i(\mathbf{W}_{i}\mathbf{z}^{(l)} + \mathbf{b}_{i}^{(l)}).
\end{equation}
Here, $\mathbf{W}_i \in \mathbb{R}^{d_{i} \times d_{l}}$ and $\mathbf{b}_i^{(l)} \in \mathbb{R}^{d_i}$ denote the weight matrix and bias vector, respectively, where $d_{l}$ is the dimension of the input $\mathbf{z}^{(l)}$ and $d_{i}$ is the dimension of the output for layer $i$, with $i = 1, \ldots, L$. The activation function $\sigma$ is applied element-wise to introduce nonlinearity into the network. Common choices for $\sigma$ include the rectified linear unit (ReLU), the sigmoid, and the hyperbolic tangent (tanh) functions. 

The Universal Approximation Theorem (UAT) provides a theoretical guarantee that a FFNN with a single hidden layer containing a finite number of neurons can approximate any continuous function on compact subsets of $\mathbb{R}^n$~\citep{cybenko_1989}. This powerful result underpins the theoretical foundation of neural networks as universal function approximators. However, the UAT does not prescribe a method for finding optimal parameters. As such, practitioners typically employ first-order gradient-based optimisation algorithms, most notably stochastic gradient descent (SGD) and its widely used variants, including Nesterov accelerated gradient (NAG)~\citep{nesterov1983method}, RMSProp, Adam~\citep{adam_opt} and AdamW~\citep{loshchilov2017decoupled}, to train these networks. These optimisation algorithms rely on the backpropagation algorithm, which efficiently computes the gradients across the network's layers by applying the chain rule. This enables effective parameter optimisation in order to minimise a chosen loss function.

\subsection{Continuous-Depth Architectures}\label{sec:continuous_depth}
On the other hand, continuous-depth architectures represent a class of models where the concept of depth is extended from a sequence of discrete transformations to a continuously parameterised process~\citep{chen_2018,massaroli_2020, queiruga2020continuous}. Instead of composing a discrete number of layers, these approaches learn how a system's state evolves smoothly over an abstract depth or a temporal variable. This framework is inherently well suited for capturing the dynamics of physical systems, where evolution is continuous, and for handling sparsely sampled data, which discrete models often struggle to represent effectively.

\subsubsection{Neural Ordinary Differential Equations}
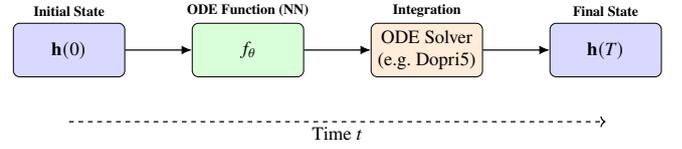
\begin{figure}
    \centering
    \resizebox{\columnwidth}{!}{%
    \begin{tikzpicture}[
      font=\Large, % �� change this to \Large or \normalsize etc. as needed
      node distance=2cm and 1.5cm,
      box/.style={draw, rounded corners, minimum height=1.2cm, minimum width=2.5cm, align=center},
      arrow/.style={-{Latex[width=2mm]}, thick},
      time/.style={font=\normalsize\bfseries, inner sep=1pt}
      ]

      % Nodes
      \node[box, fill=blue!15] (input) {\( \mathbf{h}(0) \)};
      \node[box, fill=green!15, right=of input] (func) {\( f_\theta\)};
      \node[box, fill=orange!15, right=of func] (integrate) {ODE Solver \\ (e.g. Dopri5)};
      \node[box, fill=blue!15, right=of integrate] (output) {\( \mathbf{h}(T) \)};

      % Arrows
      \draw[arrow] (input) -- (func);
      \draw[arrow] (func) -- (integrate);
      \draw[arrow] (integrate) -- (output);

      % Optional annotations
      \node[time, above=0.1cm of input] {Initial State};
      \node[time, above=0.1cm of func] {ODE Function (NN)};
      \node[time, above=0.1cm of integrate] {Integration};
      \node[time, above=0.1cm of output] {Final State};

      % Optional: time axis
      \draw[dashed, thick, ->] ($(input.south)+(0,-1)$) -- node[below] {Time $t$} ($(output.south)+(0,-1)$);

    \end{tikzpicture}
    }
    \caption{NODE computational flow: the initial state \( \mathbf{h}(0) \) is evolved forward in time via a learnt vector field \( f_\theta \), integrated using an ODE solver to produce the final state \( \mathbf{h}(T) \). Here, $t$ denotes a continuous variable that may represent either an auxiliary depth parameter or the real physical time, depending on the context.}
    \label{fig:neural_ode_architecture}
\end{figure}

Neural ordinary differential equations (NODEs)~\citep{chen_2018} are a class of continuous-depth models that can be conceptualised as continuous analogues of the widely adopted Residual Networks \footnote{Strictly speaking, the original NODE formulation does not correspond to the true deep limits of ResNets; see~\citep{dupont_2019,massaroli_2020} for more details.} (ResNets)~\citep{he_2016}. In a typical ResNet architecture, the hidden states, which are the intermediate representations of the neural network, undergoes a series of discrete transformations:
\begin{equation}\label{eq:euler_discretisation}
    \mathbf{h}_{t+1} = \mathbf{h}_{t} + f_{\boldsymbol{\theta}}(\mathbf{h}_{t}),
\end{equation}
where $\mathbf{h}_{t} \in \mathbb{R}^{d} $ represents the hidden state at layer $t$ and $f_{\boldsymbol{\theta}}: \mathbb{R}^{d} \to \mathbb{R}^{d} $ is a differentiable function parameterised by parameters $\theta$. Equation~\eqref{eq:euler_discretisation} corresponds to the forward Euler discretisation of a continuous-time dynamical system with $\Delta t = 1$. Taking the limit $\Delta t \to 0$ yields:
\begin{equation}\label{eq:neural_ode}
    \lim_{\Delta t\to0} \frac{\mathbf{h}_{t+\Delta t} - \mathbf{h}_{t}}{\Delta t} = \frac{d\mathbf{h(t)}}{dt} = f_{\boldsymbol{\theta}}(\mathbf{h}_{t}).
\end{equation}
In this formulation, dynamics of the hidden state are governed by an ordinary differential equation (ODE) parameterised by a neural network. The function $f_{\boldsymbol{\theta}}$ denotes a learnable vector field, with FFNN serving as the most canonical implementation. By specifying the initial condition $\mathbf{h}(0)$ as the input layer and interpreting $\mathbf{h}(T)$ as the output layer, the problem effectively reduces to solving an initial value problem. This can be achieved using a numerical ODE solver, which propagates the hidden states according to the following integral equation (as illustrated in Fig.~\ref{fig:neural_ode_architecture}):
\begin{equation}
	\mathbf{h}(T) = \mathbf{h}(0) + \int_{0}^{T} f_{\boldsymbol{\theta}}(\mathbf{h}_{t}) dt.
\end{equation} 
For our work, we restrict ourselves to the autonomous case, where the vector field $f_{\boldsymbol{\theta}}$ depends only on the current hidden state, i.e., $f_{\boldsymbol{\theta}}(\mathbf{h}_{t})$, and the parameters $\boldsymbol{\theta}$ are time-independent. This simplifying assumption reduces the ODE to be invariant under time translation. For the more general case of non-autonomous NODEs, where the vector field is explicitly time-dependent, see~\citep{chen_2018, dupont_2019, davis2020time} for more details.

Training NODEs can be accomplished through two primary methods: conventional backpropagation or the adjoint sensitivity method. The backpropagation approach involves storing intermediate states during the forward pass and using the chain rule to compute gradients, which can be memory-intensive for complex ODEs. Alternatively, the adjoint sensitivity method, as proposed in \citet{chen_2018}, offers a memory-efficient solution by solving a backward ODE to compute gradients. This method circumvents the need to store intermediate states, making it particularly advantageous for training deep and complex NODEs.

\section{Methodology}
This section is organised as follows. First, the dynamical quantities used to characterise the BH pair and its orbital evolution are defined. Second, a single, representative $N$-body simulation of a galaxy merger is employed as a benchmark time series for assessing the forecasting performance of classical and neural models, including a vanilla neural ordinary differential equation (NODE), for the binary evolution. Finally,we consider an ensemble of 
$N$-body simulations of galaxy mergers spanning a two-dimensional parameter space in the initial galactic orbital eccentricity and particle resolution $(e_i, N)$, within which a parameterised neural ordinary differential equation (PNODE) is constructed to learn a single unified vector field describing the BH pairs' orbital evolution conditioned on the simulation parameters.

\subsection{Dynamical descriptors of the BH pair}\label{sec:orbital_quantities}
Throughout this work we describe the BH pair using standard two-body dynamical quantities. Let $\mathbf{r}$ and $\mathbf{v}$ denote the relative separation and velocity vectors of the two BHs, and define:
\begin{equation}
    \mathbf{h} = \mathbf{r} \times \mathbf{v}, \quad h = \lVert \mathbf{h} \rVert,
\end{equation}
as the specific angular momentum vector and its magnitude. The gravitational parameter of the pair is:
\begin{equation}
    \mu = G(M_1 + M_2),
\end{equation}
where $M_1$ and $M_2$ are the BH masses.

The specific orbital energy is:
\begin{equation}
    \varepsilon = \frac{1}{2}\,\lVert \mathbf{v} \rVert^2 - \frac{\mu}{\lVert \mathbf{r} \rVert},
\end{equation}
and in the Keplerian limit, $(\varepsilon,h)$ are related to the semi-major axis $a$ and eccentricity $e$ via:
\begin{equation}
    a = -\frac{\mu}{2\varepsilon}, \quad
    e = \sqrt{1 + \frac{2\varepsilon h^2}{\mu^2}},
\end{equation}
or, equivalently,
\begin{equation}
    \varepsilon = -\frac{\mu}{2a}, \quad
    h^2 = \mu a (1-e^2).
\end{equation}
These relations allow us to switch interchangeably between $(a,e)$ and $(\varepsilon,h)$ depending on the task. In the ideal two-body problem these quantities are conserved and have the usual interpretation as orbital elements or equivalent orbital descriptors. In our simulations, however, the BHs move in a time-dependent potential, so $(a,e,\varepsilon,h)$ vary with time and should be regarded as instantaneous Keplerian approximations to the BH relative orbit. They remain highly useful as compact dynamical descriptors: $(a,e)$ track how the effective size and shape of the orbit evolve under perturbations, while $(\varepsilon,h)$ encode its binding state and kinematics, with $\varepsilon$ determining whether the BH pair is bound $(\varepsilon<0)$ or unbound $(\varepsilon>0)$, and how hard the system is, and $h$ setting the specific angular momentum, which controls how radial versus tangential the relative motion is at fixed energy.

\subsection{Emulating a Single \texorpdfstring{$N$}{N}-body Simulation}\label{sec:emulate_single_n_body}
We consider a $N$-body simulation of the merger of two equal-mass massive elliptical galaxies hosting central SMBHs, resulting in the formation of a massive BHB whose GW emission could contribute to the stochastic background detectable by Pulsar Timing Arrays. Each galaxy consists of a stellar bulge, a dark matter halo, and a central SMBH. We set $\mbh = 5.0\times10^8\msun$ and $\mbulge = 10^{11}\msun$, consistent with scaling relations for progenitors of PTA sources~\citep{Reines2015}. The bulge follows a Sérsic profile~\citep{sersic1963,sersic1968}, and the halo follows an NFW profile~\citep{NFW1996}, with initial conditions generated via the AGAMA software~\citep{agama_vasiliev_2019} and a central mass refinement scheme~\citep{attard2024}.

The galaxies are placed initially at $100\kpc$ on a Keplerian orbit with eccentricity $e_0 = 0.9$ and evolved using the $N$-body code {\tt Griffin}~\citep{dehnen_2014} until the BHB reaches separations of $\sim$pc. Using the dynamical descriptors defined in Section~\ref{sec:orbital_quantities}, we compute the semi-major axis $a$ and eccentricity $e$ of the BH binary from the relative phase-space coordinates at each snapshot. From 673 raw snapshots, we select 210 corresponding to a stable bound binary, yielding a short time series:
\begin{equation}
    \mathcal{D} = \{(t_i, \mathbf{q}_i)\}_{i=1}^{N}, \quad \mathbf{q}_i = (\ln(a)_i, e_i),
\end{equation}
which we use to benchmark classical and neural forecasting models. Detailed model specifications, data preprocessing, and training procedures for this single-simulation experiment are provided in Appendix~\ref{app:single_sim}.

\subsection{Emulating an Ensemble of \texorpdfstring{$N$}{N}-body Simulations}
We extend the NODE framework to a broader suite of galaxy merger $N$-body simulations. All simulations are generated from the same underlying $N$-body model but span a two-dimensional parameter space $\mathcal{P} = \{(e_i, N) : e_i \in \mathcal{E}, N \in \mathcal{N}\}$, here $\mathcal{E} = \{0.97, 0.9, 0.99\}$ and $\mathcal{N} = \{1\text{M}, 2\text{M}, 4\text{M}, 8\text{M}, 16\text{M}, 32\text{M}, 64\text{M}\}$ represent initial galactic orbital eccentricities and particle resolutions (where $1\text{M}=2^{20}=1048576$), respectively. Independent random realisations are generated for each $(e_{i}, N)$, yielding 156 distinct BH phase-space trajectories with $\sim200$ timesteps per trajectory. Although the simulations also include stellar particles, we focus exclusively on the BH trajectories in this work.

Each progenitor is modelled as a single-component stellar bulge following a Dehnen profile~\citep{dehnen1993family}:
\begin{equation}
    \rho(r) = \frac{(3-\gamma)M}{4\pi} \frac{r_0}{r^{\gamma} (r + r_0)^{4-\gamma}},
\end{equation}
with total stellar mass \(M\), scale radius \(r_0\) and inner slope \(\gamma\). We adopt shallow cusps with \(\gamma = 0.5\), as in \citet{nasim2020}, and assign to each galaxy a central BH of mass \(M_{\bullet} = 0.0025\,M\). All simulations are evolved in dimensionless units with \(M_{\rm tot} = G = r_0 = 1\), where \(M_{\rm tot}\) is the combined mass of the two progenitors. For the physical interpretation of merger timescales in this work, we adopt a physical scaling in which the central BH mass corresponds to \(M_{\bullet} = 10^9\,{\rm M_\odot}\); other PTA-relevant masses can be obtained a posteriori using the BH influence radius together with the \(M_{\bullet}\)–\(\sigma\) relation of \citet{kormendy_coevolution_2013}.

To obtain a compact description of the BH dynamics across the ensemble, we work with the specific orbital energy \(\varepsilon\) and angular momentum \(h\) defined in Section~\ref{sec:orbital_quantities}. In the single-simulation baseline of Section~\ref{sec:emulate_single_n_body}, we forecast the Keplerian orbital elements \((\ln(a), e)\) directly, as they provide an intuitive description of the BHB's orbit. For the ensemble of simulations, however, \((a,e)\) are well defined only once the binary is bound, which significantly reduces the amount of usable training data, and across extended time intervals their evolution is approximately linear, providing limited nonlinear structure and hence weaker dynamical distinguishability between trajectories. Alternatively, the corresponding trajectories in \((\varepsilon, h)\) display richer secular and short-timescale variations throughout the evolution, with better dynamical distinguishability for a PNODE to learn. We therefore model the BH dynamics in terms of \((\varepsilon, h)\) and recover the corresponding \((a,e)\) a posteriori using the relations in Section~\ref{sec:orbital_quantities}. Now the NODE is parameterised by, and explicitly conditioned on, the simulation parameters \((e_i, N) \in \mathcal{P}\), and thus defines a parameterised NODE (PNODE) of the form:
\begin{equation}
    \frac{d\mathbf{q}(t)}{dt}
= f_\theta\big(\mathbf{q}(t);\, e_i, N\big),
\end{equation}
where $\mathbf{q}(t)=(\varepsilon(t), h(t))$. As such, a single PNODE model learns the joint evolution across the simulation, while conditioning on simulation parameters $(e_i, N) \in \mathcal{P}$, enabling the model to flexibly adapt its learned dynamics to each region of parameter space, rather than fitting a generic, averaged vector field across the BH trajectories. 

For training, we standardised $\varepsilon$ and $h$ component-wise over the full simulation suite, and rescaled each simulation time axis to $t\in[0,1]$. The set of 156 BH trajectories are partitioned into training, validation and test subsets using an 75/12.5/12.5 split at the trajectory level, stratified by $(e_i, N)$ to ensure each subset contains proportional representation of each parameter configuration across all splits. Our implementation of the PNODE is a fully connected feedforward network with four hidden layers of width 64 and the ELU (exponential linear unit) activations, mapping $(\mathbf{q}(t), e_i,N)$ to the time derivative $d\mathbf{q}(t)/dt$. During both training and evaluation, we integrate the PNODE with an adaptive Dormand-prince (Dopri5) solver with relative tolerance $\text{rtol}=10^{-7}$ and absolute tolerance $\text{atol}=10^{-8}$. Training uses the AdamW optimiser with a two-stage curriculum approach: in the first stage, the model is trained by exposing only the first half of each trajectory with a learning rate of $2\times10^{-3}$, then it is fine-tuned on the full trajectories in the second stage with a slightly lower learning rate of $2\times10^{-3}$. Throughout both stages, we optimised a weighted Huber loss:
\begin{equation}
    \mathcal{L} = w_{\varepsilon} \mathcal{L}_{\varepsilon} + w_{h} \mathcal{L}_{h},
\end{equation}
where $\mathcal{L}_{\varepsilon}$ and $\mathcal{L}_{h}$ are the Huber losses (with transition parameter $\delta=0.1$) on the standardised $\varepsilon$ and $h$ respectively. $w_{\varepsilon}$ and $w_{h}$ are weighting factors, and we set $w_{\varepsilon}=1$ and $w_{h}=8$ to up-weight accuracy in the angular momentum component, since the determination  of eccentricity is particularly sensitive to small variations in $h$ when the binary is bound. Additionally, the OneCycleLR learning rate scheduler and early stopping based on the validation loss was used to improve training stability and to prevent overfitting.

\section{Results} \label{sec:results}
This section is organised as follows. We first present a baseline forecasting experiment on a single \(N\)-body simulation, comparing several classical and neural models for predicting the Keplerian orbital elements, and summarise the main findings while referring readers to Appendix~\ref{app:single_sim} for full details. We next consider the ensemble of \(N\)-body simulations and assess the performance of the PNODE in recovering the evolution of the specific orbital energy and angular momentum, as well as the corresponding Keplerian orbital elements for a held-out test set.

\subsection{Single \texorpdfstring{$N$}{N}-body simulation Forecasting Baseline}
We first consider a baseline task: forecasting the Keplerian orbital elements \((\ln(a), e)\) for a single \(N\)-body simulation using several classical and neural models (FFNN, BNN, LSTM, XGBoost, VARIMA and NODE). Across different train/test splits, all methods attain comparable mean absolute errors, with no evidence that any one model consistently dominates. Qualitative differences (e.g. trajectory smoothness or small discontinuities at the train--test boundary) mainly reflect differing inductive biases and forecasting strategies rather than a robust performance hierarchy. For this approximately linear forecasting interval, strongly regularised models such as the BNN perform at least as well as the NODE, suggesting limited benefit from the added flexibility of using a NODE in this regime. Detailed numerical results and forecast plots are provided in Appendix~\ref{app:single_sim}. In the remainder of this section, we therefore focus on the more realistic setting, where a PNODE is trained across an ensemble of \(N\)-body simulations.

\subsection{PNODE on Ensemble \texorpdfstring{$N$}{N}-body Simulations} \label{sec:pnode_ensemble_results}
\begin{figure*}
    \centering
    \begin{subfigure}{0.33\textwidth}
        \centering
        \includegraphics[width=\linewidth]{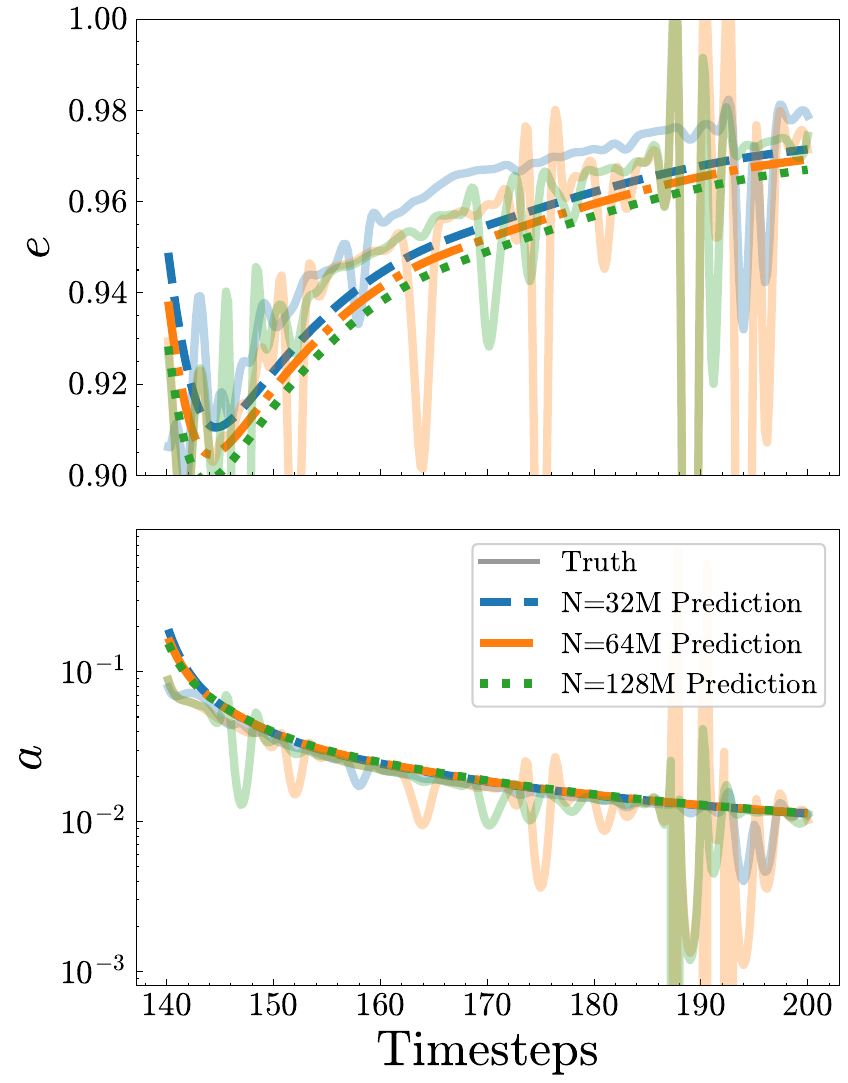}
        \caption{$e_i=0.9$}
        \label{fig:ecc09}
    \end{subfigure}\hfill
    \begin{subfigure}{0.33\textwidth}
        \centering
        \includegraphics[width=\linewidth]{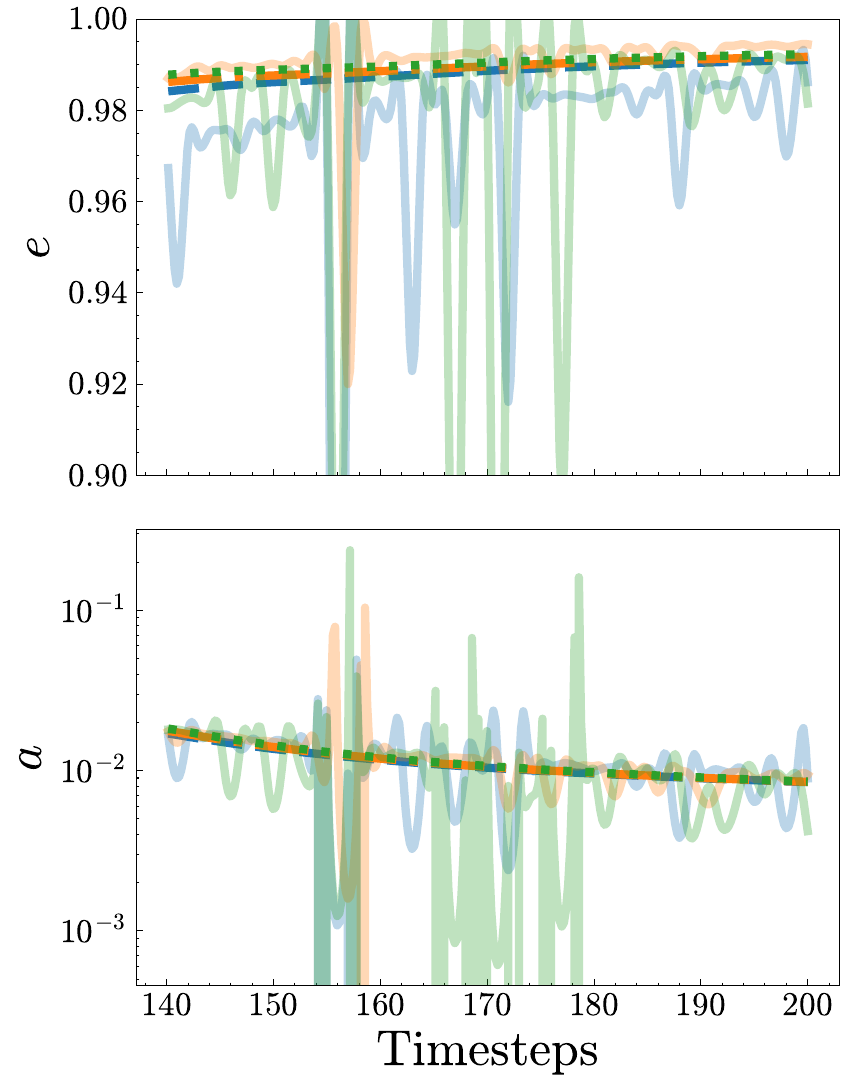}
        \caption{$e_i=0.97$}
        \label{fig:ecc097}
    \end{subfigure}\hfill
    \begin{subfigure}{0.33\textwidth}
        \centering
        \includegraphics[width=\linewidth]{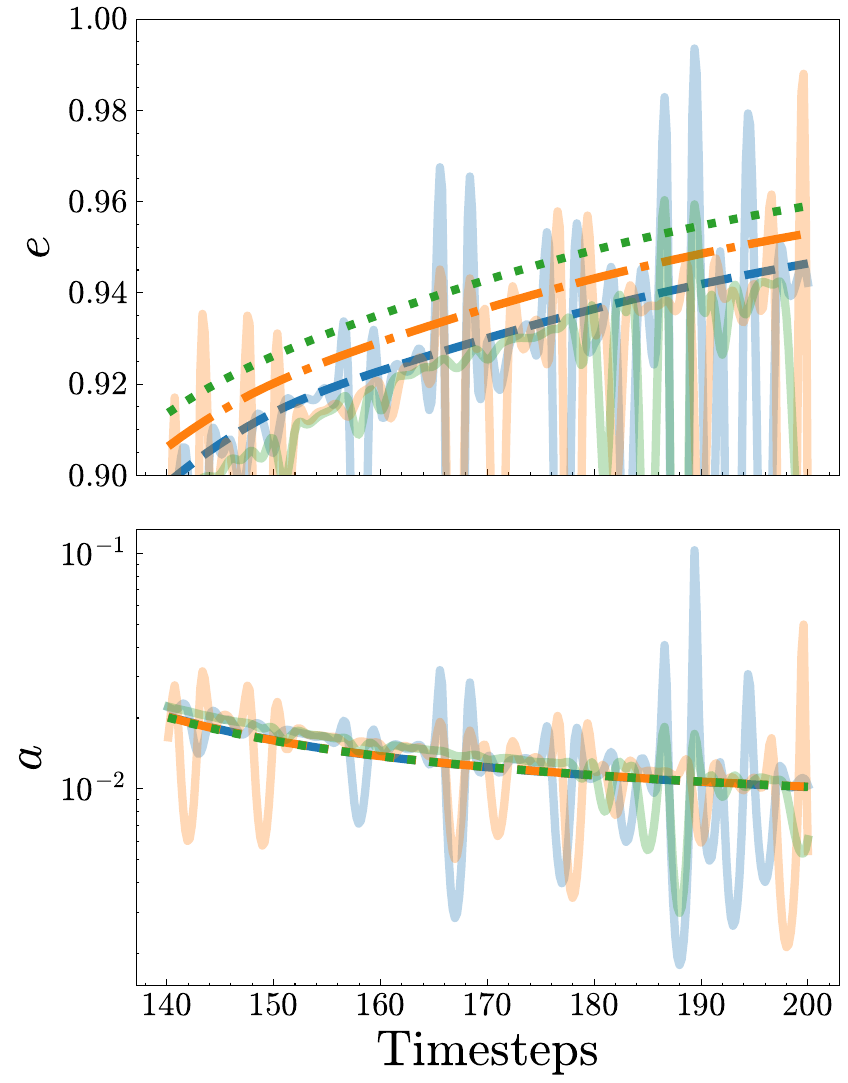}
        \caption{$e_i=0.99$}
        \label{fig:ecc099}
    \end{subfigure}
    \caption{Predictions of PNODE for $e$ and $a$ (dashed lines) against truth (simulation) for selected BH trajectories in test set for initial galactic orbital eccentricities $e_{i}=0.9, 0.97, 0.99$ at resolutions (32M, 64M, 128M). All quantities are shown in the dimensionless code units of the simulations.}
    \label{fig:pnode_predictions}
\end{figure*}

The results for selected BH trajectories for $(e_i, N)$ are shown in Fig. \ref{fig:pnode_predictions}. Although the held out test set includes 24 distinct trajectories, we display only a single representative realisation for the resolutions with the sparsest coverage in the dataset (32M, 64M, and 128M). The test set is in-distribution with respect to most of the $(e_i, N)$ parameter space, but partially out-of-distribution (OOD) in resolution: in particular, no 128M simulations are included in the training set, and only a single realisation is available in the test set. For each case, the trained PNODE is integrated forward in time from the corresponding initial conditions to generate the predicted trajectories for $\varepsilon$ and $h$. From Fig. \ref{fig:pnode_predictions}, we see that the PNODE predictions closely track the actual values of $\varepsilon$ and $h$ across initial galactic orbital eccentricities $e_{i}=0.9, 0.97, 0.99$ at resolutions (32M, 64M, 128M), capturing the overall secular evolution of the BHs. Throughout this section, $\varepsilon$, $h$, $a$ and $e$ are expressed in the dimensionless units of the $N$-body simulations, before any physical scaling is applied. Across the 24 held-out trajectories, the PNODE attains median absolute errors of $1.3\times10^{-5}$ in $\varepsilon$ and $6.7\times10^{-6}$ in $h$, corresponding to fractional errors of $\sim10^{-2}$ ($1\%$) relative to their typical dynamical ranges. Particularly, the model is able to reproduce the broad initial peak in the specific orbital energy $\varepsilon$ associated with the first pericentre passage, where large-scale torques drive a rapid rearrangement of the BH orbital energy. However, the PNODE under-resolves the sharper, irregular changes in energy and angular momentum driven by subsequent pericentre passages and strong three-body encounters, while accurately capturing the smoother secular evolution. This behaviour is qualitatively consistent with the tendency of deep neural networks to favour smooth, low-frequency structure (spectral bias;~\citep{rahaman2019spectral, xu2025overview}), and may also be reinforced by the continuity bias of NODEs and the choice of loss function and other implicit regularisation mechanisms.

Overall, these results suggest that for this ensemble, a single PNODE conditioned on simulation parameters $(e_i, N)$ is able to learn a unified vector field that interpolates smoothly across the sampled region of parameter space and shows modest resolution-level extrapolation. We note that the training set is dominated by lower-resolution simulations, so the relatively robust performance at $N=128\text{M}$ should not be over-interpreted as evidence of substantial extrapolative ability: the $N=128\text{M}$ case remains dynamically similar to $N=32\text{M}, 64\text{M}$ and conditioning on $(e_i, N)$ allows the PNODE to vary its dynamics smoothly as a function of $N$ for a given $e_i$, capturing shared global dynamics with resolution-specific biases rather than collapsing to a low-resolution dominated solution.

\subsubsection{Recovery of orbital elements}
\begin{figure*}
    \centering
    \begin{subfigure}{0.33\textwidth}
        \centering
        \includegraphics[width=\linewidth]{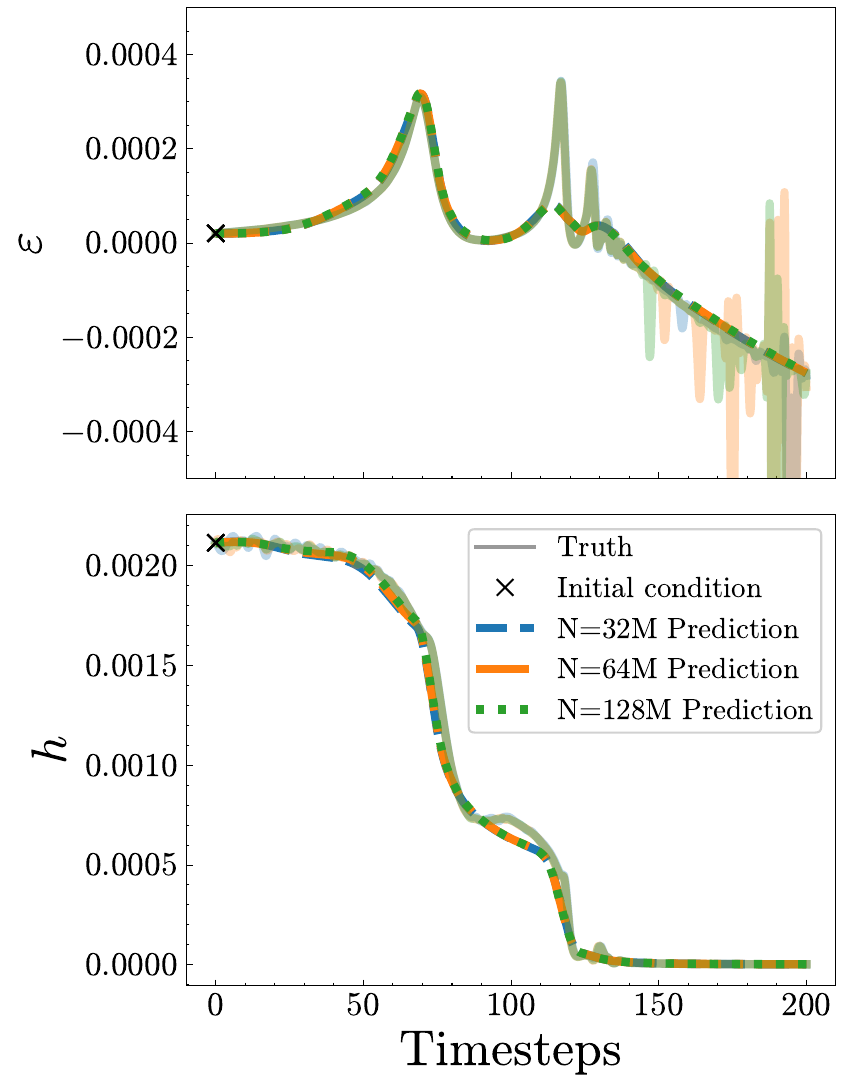}
        \caption{$e_i=0.9$}
        \label{fig:ecc09}
    \end{subfigure}\hfill
    \begin{subfigure}{0.33\textwidth}
        \centering
        \includegraphics[width=\linewidth]{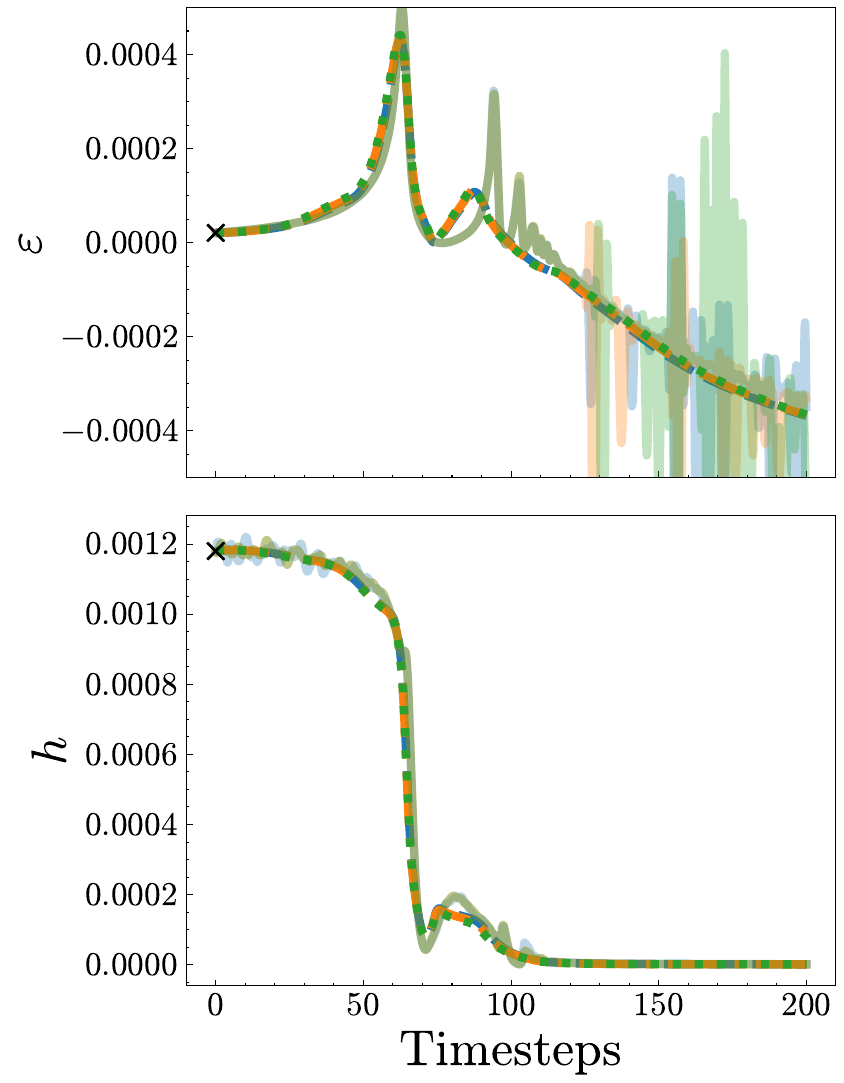}
        \caption{$e_i=0.97$}
        \label{fig:ecc097}
    \end{subfigure}\hfill
    \begin{subfigure}{0.33\textwidth}
        \centering
        \includegraphics[width=\linewidth]{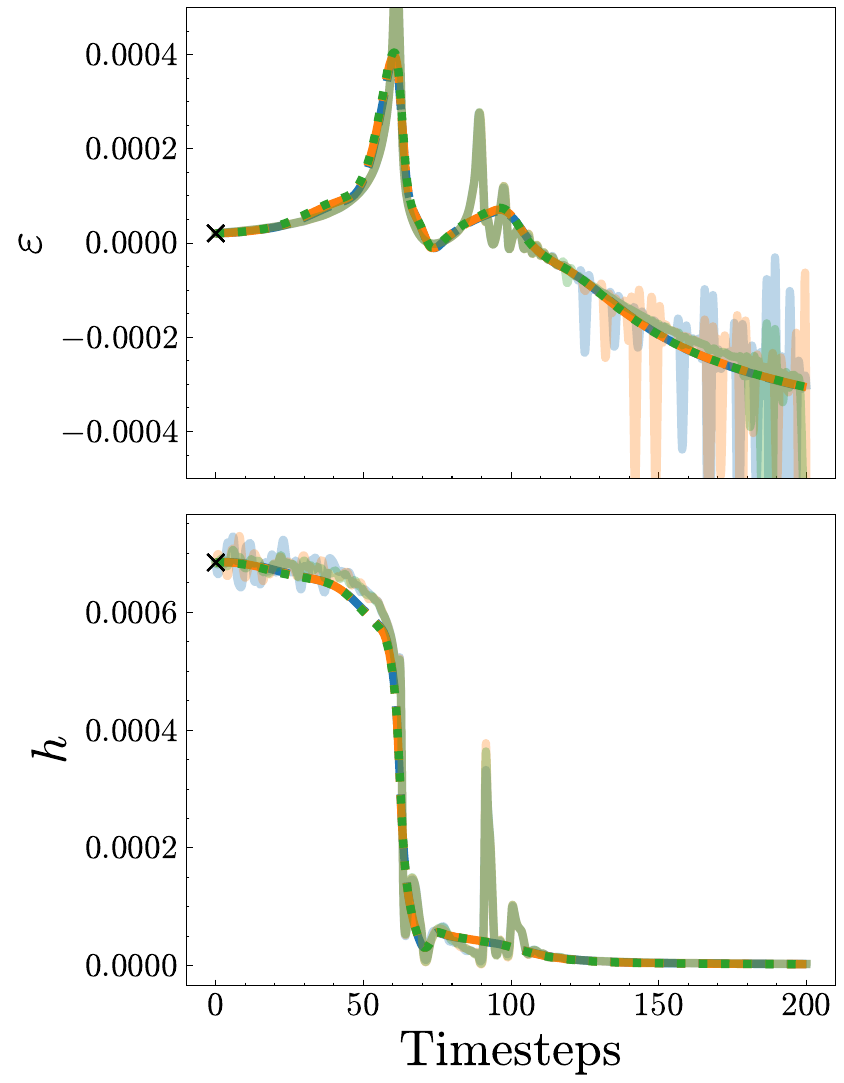}
        \caption{$e_i=0.99$}
        \label{fig:ecc099}
    \end{subfigure}
    \caption{Predictions of PNODE for $\varepsilon$ and $h$ (dashed lines) against truth (simulation) for selected BH trajectories in test set for initial galactic orbital eccentricities $e_{i}=0.9, 0.97, 0.99$ at resolutions (32M, 64M, 128M). All quantities are shown in the dimensionless code units of the simulations.}
    \label{fig:pnode_predictions_orb_elem}
\end{figure*}

For each trajectory in the test set, once the BH pair has hardened and remains bound \((\varepsilon < 0)\), we map the PNODE predictions for $(\varepsilon, h)$ back to the Keplerian orbital elements $(a,e)$ using the relations defined in Section~\ref{sec:orbital_quantities}. Fig.~\ref{fig:pnode_predictions_orb_elem} compares PNODE predictions for \(e\) and \(a\) (dashed) to the \(N\)-body results (solid) for the selected test trajectories. The instantaneous orbital elements derived from \((\varepsilon, h)\) contain pronounced short-timescale spikes, particularly near pericentre passages, reflecting the high sensitivity of \(a(t)\) and \(e(t)\) to small, rapidly varying perturbations in energy and angular momentum. From Fig.~\ref{fig:pnode_predictions_orb_elem}, the PNODE recovers the long-term evolution of the semi-major axis across all \((e_i, N)\) with only modest discrepancies, while for eccentricity the same smoothing generally captures the broad global trend but, in the most radial case \(e_i = 0.99\), leads to more noticeable deviations than in the other configurations. Since the underlying secular evolution of the orbital elements is expected to be smooth, we train the PNODE directly on the raw trajectories and rely on its smooth vector field to average over high-frequency fluctuations. In principle, we could remove the noisiest pericentre points, but this would substantially reduce the effective size and phase-space coverage of the training set, so we retain them.

\section{Merger timescales}
\label{sec:timescales}
Theoretical estimates of BHB merger timescales are essential for predicting the stochastic GW background measured by PTAs and for future detections with LISA. However, $N$-body simulations of BHB formation and evolution in galactic mergers are computationally expensive even with state-of-the-art codes such as {\sc Griffin}~\citep{dehnen_2014} and {\sc KETJU}~\citep{rantala_2017}, and are therefore generally limited to the first two evolutionary phases, typically down to at most the hard-binary separation. The eccentricity of the BHB at formation is a key quantity setting the long-term evolution from binary formation to coalescence. \citet{gualandris2022} show that it is primarily set by the orbital eccentricity of the galaxies, but finite resolution introduces significant scatter around this relation. Numerical stochasticity affects the binary eccentricity much more strongly than its semi-major axis, as eccentricity is more susceptible to perturbations. \citet{nasim2020} find that the dispersion in eccentricity measured over a number of random realisations of the same models scales with resolution (total particle number $N$ within a characteristic radius) approximately as $1/\sqrt{N}$, and particle numbers in excess of $10^{7}$ are required for a $\sim 10\%$ error on the merger timescale, with nearly radial mergers showing the largest variance~\citep{fastidio2024}. These considerations highlight that merger-time predictions are highly sensitive to the eccentricity and orbital state at binary formation, and that these quantities are difficult to determine robustly even in state-of-the-art $N$-body simulations.

The further evolution of the BHB, beyond the hard-binary regime, can be efficiently treated with semi-analytical models (SAMs) that include the effects of stellar hardening and GW emission. Such models are quick to run and have been successfully used to bridge the gap between the end of the $N$-body simulation and the final coalescence of the MBHs, providing estimates of the total time to coalescence~\citep[e.g.][]{gualandris_2012,attard2024,fastidio2024}. SAMs are usually calibrated either on scattering experiments~\citep[e.g.][]{fastidio2024} or directly on $N$-body simulations~\citep[e.g.][]{nasim2020,attard2024}, where the rate of change of the orbital elements is inferred from the simulation data and extrapolated to late times via standard numerical methods. We employ the SAM described in \citet{attard2024} to model the evolution of the BHB from the end of the $N$-body simulation (or somewhat earlier) through the final phase of stellar hardening and GW emission until the two MBHs coalesce, thereby estimating the total merger timescale. The BHB is in the loss-cone refilling phase when the SAM commences, with hardening initially driven by interactions with background stars; if the BHB continues to shrink efficiently, GW emission takes over and drives the final inspiral and coalescence, aided by the triaxiality of the merger remnant and the associated repopulation of the loss cone~\citep{vasiliev_2015,gualandris_2017}.

The model evolves the semi-major axis and eccentricity according to the coupled differential equations:
\begin{align}
    \frac{da}{dt} &= \left.\frac{da}{dt}\right|_{*} + \left.\frac{da}{dt}\right|_{\rm GW}, \label{eq:dadt} \\
    \frac{de}{dt} &= \left.\frac{de}{dt}\right|_{*} + \left.\frac{de}{dt}\right|_{\rm GW}, \label{eq:dedt}
\end{align}
where the stellar-hardening contributions are written as:
\begin{align}
    \left.\frac{da}{dt}\right|_{*} &= -s(t)\,a^{2}, \label{eq:dadt_stars} \\
    \left.\frac{de}{dt}\right|_{*} &= s(t)\,K\,a, \label{eq:dedt_stars}
\end{align}
with the time-dependent hardening rate:
\begin{equation}
\label{eq:hard_rate} 
s(t)  = \frac{d}{dt}\left(\frac{1}{a}\right)
\end{equation}
and the eccentricity growth rate:
\begin{equation}
\label{eq:k_rate} 
K  = \frac{de}{d\ln(1/a)}\,.
\end{equation}
The GW terms follow the standard formulae for two point masses $M_1$ and $M_2$~\citep{peters1964}:
\begin{equation}
    \left.\frac{da}{dt}\right|_{\rm GW} = -\frac{64}{5}\frac{G^{3}M_{1}M_{2}M_{T}}{c^{5}a^{3}(1-e^{2})^{7/2}}
    \left( 1+\frac{73}{24}e^{2}+\frac{37}{96}e^{4} \right),
	\label{eq:peters_a}
\end{equation}
\begin{equation}
    \left.\frac{de}{dt}\right|_{\rm GW} = -\frac{304}{15}e\frac{G^{3}M_{1}M_{2}M_{T}}{c^{5}a^{4}(1-e^{2})^{5/2}}
    \left( 1+\frac{121}{304}e^{2} \right),
	\label{eq:peters_e}
\end{equation}
where $M_{T}=M_{1}+M_{2}$ is the total binary mass. These equations highlight the strong dependence of the orbital evolution on eccentricity, making the eccentricity and orbital elements at the SAM handoff time a key source of uncertainty. The SAM solves the system of coupled ODEs given in equations~\eqref{eq:dadt} and~\eqref{eq:dedt}, starting from initial conditions for the semi-major axis and eccentricity and requiring, in addition, the hardening rate and eccentricity growth rate as defined in equations~\eqref{eq:hard_rate} and~\eqref{eq:k_rate}. These initial conditions and rates can be inferred directly from $N$-body data when sufficiently long, high-resolution simulations are available, but in practice they are often limited by numerical noise and resolution. This motivates the use of surrogate models, such as a parameterised NODE (PNODE) trained on an ensemble of $N$-body simulations spanning a region of simulation parameters, which can provide smoother estimates of the hardening and eccentricity growth rates and reconstruct smooth orbital elements across the sampled parameter space. Moreover, when higher-resolution runs are dynamically similar to lower-resolution ones, such surrogates can provide reasonable estimates for orbital elements even for modestly out-of-sample resolutions, before the quantities are passed to the semi-analytical models for BHB merger timescale calculations.

In practice, the time-dependent hardening rate $s(t)$ and eccentricity growth rate $K$ are estimated directly from the orbital elements using finite-difference approximations. Starting from the time of binary formation, we obtain $a$ and $e$ from either the $N$-body simulations or the PNODE-predicted values, and evaluate $s(t) \approx \Delta(1/a)\Delta t$ and $K \approx \Delta e / \Delta\ln(1/a)$ over small time intervals. The resulting estimates of $s(t)$ and $K$ are then fitted with simple exponential functions, which are used as inputs to the SAM to provide the late-time extrapolation of the evolution required to compute merger timescales.

\begin{figure*}
	\centering
	\includegraphics[width=1\textwidth]{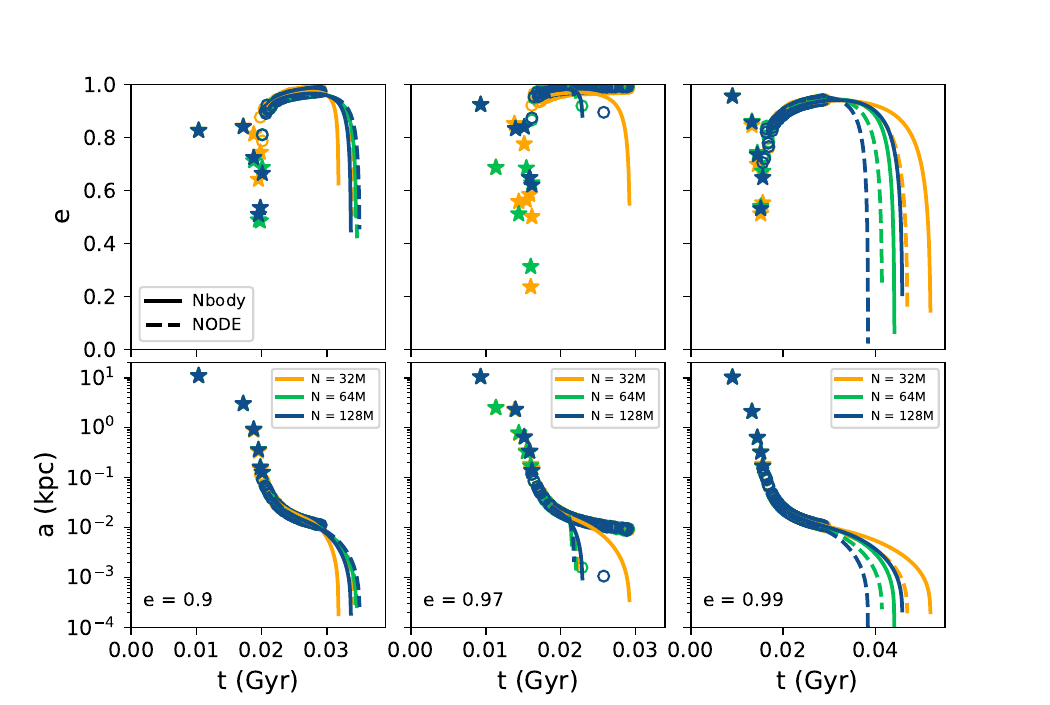}
    \caption{Evolution of the orbital elements $a$ and $e$ of the BHB as a function of time, from the start of the $N$-body simulation to coalescence of the MBHs. The lines represent the elements calculated via the semi-analytical model, with hardening rates $s$ and $K$ from a direct numerical differentiation and fit of both the $N$-body data (solid lines) and the predictions from PNODE (dashed lines).  The circles represent the original $N$-body data, , from the time of binary formation to the end of the simulation, while the stars represent the orbital elements calculated at early times, when the BHs are still unbound, from the apocentre and pericentre values. The total evolutionary time from the beginning of the simulation to coalescence provides an estimate of the total merger timescale of the BHBs, as predicted through extrapolation.}	\label{fig:merger_timescale_comparisons}
\end{figure*}

For the nine representative test trajectories spanning $(e_i, N)=\{0.9,0.97,0.99\} \times {32\text{M}, 64\text{M}, 128\text{M}}$, we compare merger timescales obtained when the SAM is fed with rates and orbital elements derived directly from $N$-body simulations to those obtained from PNODE-predicted orbital elements, which is shown in Fig.~\ref{fig:merger_timescale_comparisons}. We find that the absolute differences in total merger time range from $\sim 1.4\times10^{5}\,\mathrm{yr}$  to $\sim 7.4\times10^{6}\,\mathrm{yr}$, corresponding to relative differences between $\lesssim 1\%$ and $\sim 24\%$. For most cases, the discrepancy remains at the few-percent level, well within typical uncertainties from $N$-body models. We note that the largest deviations occur for the $(e_i=0.97, 32\text{M})$ and $(e_i=0.99, 128\text{M})$ models, which show a more visible departure of the NODE predicted elements from the $N$-body data (see Fig.~\ref{fig:pnode_predictions}). Given the strong dependence of the Peters' equation on the initial values, particularly on the eccentricity, such small deviations lead to more significant differences in mergers times. Even these cases, however, remain well within the uncertainties of current state-of-the-art predictions via either simulations or semi-analytical models~\citep[e.g.][]{2022Sykes, 2022IzquierdoVillalba}.

\section{Discussion and Conclusions}
\label{sec:discussion}

In this work, we have investigated when neural ordinary differential equation (NODE) models are useful for forecasting the dynamics of BHBs, and under what conditions a parameterised neural ordinary differential equation (PNODE) becomes an effective surrogate for the secular evolution of BH pairs in an ensemble of $N$-body simulations of galaxy mergers, represented in terms of their specific orbital energy and angular momentum. For a single $N$-body realisation in which the Keplerian orbital elements are approximately linear over the forecasting interval, all tested models (FFNN, BNN, LSTM, XGBoost, VARIMA, NODE) yield similar mean absolute errors, differing mainly in smoothness and behaviour at the train–test boundary. In this simple setting, a well-regularised model such as a BNN or a classical statistical baseline is sufficient, and there is no clear evidence that a NODE provides any additional advantage for forecasting a single, noisy trajectory that appears approximately linear only over the chosen forecast interval.

For the ensemble of $N$-body simulations spanning the two-dimensional parameter space of initial galactic orbital eccentricity and particle resolution $(e_i, N)$, the PNODE, conditioned on $(e_i, N)$ and trained directly on the compact set of descriptors $(\varepsilon, h)$, the PNODE learns a unified vector field that captures the secular evolution of BH pairs across the sampled region of parameter space, including partially out-of-distribution resolutions. Across the held-out trajectories in the test set, the PNODE attains median fractional errors of order $10^{-2}$ in both $\varepsilon$ and $h$, demonstrating that a single model can interpolate effectively between different $(e_i, N)$ while smoothly adjusting its dynamics with resolution rather than collapsing to a low-resolution-dominated behaviour. As already noted in Section~\ref{sec:pnode_ensemble_results}, the PNODE systematically smooths over short-timescale fluctuations in $\varepsilon$ and $h$, such as the sharp variations near pericentre passages, effectively averaging these spikes into smoother orbital elements during the hardening phase.

Mapping the PNODE-predicted $(\varepsilon, h)$ back to Keplerian elements and coupling them to a semi-analytical model (SAM) that includes stellar hardening and gravitational-wave emission, we obtain merger timescales that typically agree with those derived directly from the $N$-body elements to within a few percent, and at worst to $\sim 24\%$. These differences are comparable to the spread induced by finite resolution and other modelling choices in existing $N$-body+SAM estimates of BHB merger times, and reflect the strong sensitivity of Peters-like timescale formulae to the initial eccentricity. In this sense, the PNODE provides orbital elements at the SAM handoff time that are adequate for exploring coalescence times within our setup and, by producing smoother and more densely sampled orbital elements compared to the raw $N$-body outputs, can help improve the finite-difference estimates of the hardening and eccentricity-growth rates used in the semi-analytical model.

Within the presently sampled region of $(e_i, N)$, our results indicate that a resolution-aware PNODE can interpolate between nearby simulations without degrading merger-time estimates beyond existing theoretical uncertainties. This suggests that, given a broader and more diverse training ensemble, similar surrogates could eventually be used to explore merger timescales in sparsely sampled parts of parameter space.

Several limitations of the present work must be noted. The training ensemble is built from a single-component galaxy model and consists exclusively of equal-mass mergers with equal-mass black holes, exploring only the associated two-dimensional parameter space of initial eccentricity and particle resolution $(e_i, N)$. The demonstrated PNODE performance should therefore be regarded as specific to this particular $N$-body setup and parameter range, rather than as a general statement about other galaxy models or merger configurations. Within this framework, merger timescales are computed by passing PNODE-derived orbital elements to an external semi-analytical model, so the stellar-hardening and gravitational-wave-driven evolution remain decoupled from the PNODE itself. A more integrated and potentially more robust approach would be to let the PNODE learn the effective hardening and eccentricity-growth rates directly from the $N$-body data, and then apply semi-analytical prescriptions, together with Peters-like gravitational-wave terms, as a correction layer, rather than relying on purely data-driven extrapolation combined only with Peters’ equations. Finally, the current PNODE is deterministic and does not provide explicit uncertainty estimates on the inferred orbital evolution or merger times. A more complete treatment would incorporate uncertainty quantification and training on a broader set of $N$-body setups so that both predictive accuracy and model uncertainties can be assessed more systematically. In addition, because the PNODE is explicitly conditioned on simulation parameters, its differentiable structure could be exploited in future to perform sensitivity analyses of orbital evolution and merger times with respect to a more general set of host and simulation parameters (beyond $(e_i, N)$), helping to identify which regions of parameter space have the strongest impact on predicted merger timescales, and therefore where additional high-resolution simulations would be most informative.

\section*{Acknowledgements}
AG would like to acknowledge support from the STFC grant ST/Y002385/1. PD is supported by a UKRI Future Leaders Fellowship (grant reference MR/S032223/1)

%%%%%%%%%%%%%%%%%%%%%%%%%%%%%%%%%%%%%%%%%%%%%%%%%%
\section*{Data and Software Availability}
The code and data underlying this article will be shared on reasonable request to the corresponding authors. The NODE/PNODE is implemented using the PyTorch framework ~\citep{steiner2019pytorch}, with ODE solvers from the \texttt{torchdiffeq} library ~\citep{chen_2018}. Several baseline models were implemented using the SciPy library, and data analysis and plotting were carried out using pandas, NumPy and Matplotlib.

%%%%%%%%%%%%%%%%%%%% REFERENCES %%%%%%%%%%%%%%%%%%

% The best way to enter references is to use BibTeX:

\bibliographystyle{mnras}
\bibliography{references} % if your bibtex file is called example.bib

% Alternatively you could enter them by hand, like this:
% This method is tedious and prone to error if you have lots of references
%\begin{thebibliography}{99}
%\bibitem[\protect\citeauthoryear{Author}{2012}]{Author2012}
%Author A.~N., 2013, Journal of Improbable Astronomy, 1, 1
%\bibitem[\protect\citeauthoryear{Others}{2013}]{Others2013}
%Others S., 2012, Journal of Interesting Stuff, 17, 198
%\end{thebibliography}

%%%%%%%%%%%%%%%%%%%%%%%%%%%%%%%%%%%%%%%%%%%%%%%%%%

%%%%%%%%%%%%%%%%% APPENDICES %%%%%%%%%%%%%%%%%%%%%

\appendix
\FloatBarrier

\section{Single \texorpdfstring{$N$}{N}-body Forecasting Experiment}
\label{app:single_sim}

\subsection{Dataset construction and models}
We construct supervised forecasting pairs from the time series
$\mathcal{D} = \{(t_i, \mathbf{q}_i)\}_{i=1}^{N}$, with
$\mathbf{q}_i = (\ln(a_i), e_i)$, using a sliding window of length $w$
and an $n$-step forecast horizon:
\begin{align}
\mathbb{X}_t^{(w)} &= [\mathbf{q}_{t-w+1},\dots,\mathbf{q}_t]^\top \in \mathbb{R}^{w\cdot d},\\
\mathbb{Y}_t^{(n)} &= [\mathbf{q}_{t+1},\dots,\mathbf{q}_{t+n}]^\top \in \mathbb{R}^{n\cdot d},\quad d=2,
\end{align}
with a chronological train/test split to preserve temporal consistency.

We benchmark a set of machine-learning and classical forecasting models on this single-simulation dataset:
\begin{description}
\item[FFNN:] A feedforward neural network that maps a fixed-length input window $\mathbb{X}_t^{(w)}$ to the $n$-step forecast $\mathbb{Y}_t^{(n)}$ via a single forward pass.
\item[BNN:] The same architecture and input–output structure as the FFNN but with distributions over weights; multi-step forecasts are obtained by averaging predictions over posterior samples.
\item[NODE:] A neural ODE parameterising the time derivative of the orbital elements with a feedforward network; forecasts are generated by integrating the learned vector field from the last observed state.
\item[LSTM:] A recurrent sequence model that processes time-ordered inputs and produces multi-step forecasts recursively by feeding predictions back as inputs.
\item[VARIMA:] A vector autoregressive integrated moving-average model fitted to the full past trajectory and used in a standard recursive forecasting scheme.
\item[XGBoost:] Gradient-boosted decision trees applied in a direct multi-output configuration, using sliding windows of past states to predict all $n$ future steps simultaneously.
\end{description}

\subsection{Forecasting results}
\begin{table}
	\centering
	\caption{Mean Absolute Error (MAE) of individual models for forecasting at chronological train/test splits. The best-performing model for each split is highlighted in bold.}
	\label{tab:forecast_performance}
	\begin{tabular}{lccc}
		\toprule
		Model & 80/20 Split & 70/30 Split & 60/40 Split \\
		\midrule
		NODE    & 0.0953           & 0.1108           & 0.1213 \\
		FFNN    & 0.1366           & 0.1642           & 0.1784 \\
        BNN     & \textbf{0.0910}  & \textbf{0.1085}  & \textbf{0.0961} \\
		LSTM    & 0.1231           & 0.1314           & 0.2719 \\
		XGBoost & 0.1955           & 0.2436           & 0.2863 \\
		VARIMA  & 0.1863           & 0.2350           & 0.3487 \\
		\bottomrule
	\end{tabular}
\end{table}
\begin{figure*}
    \centering
    \includegraphics[width=\textwidth]{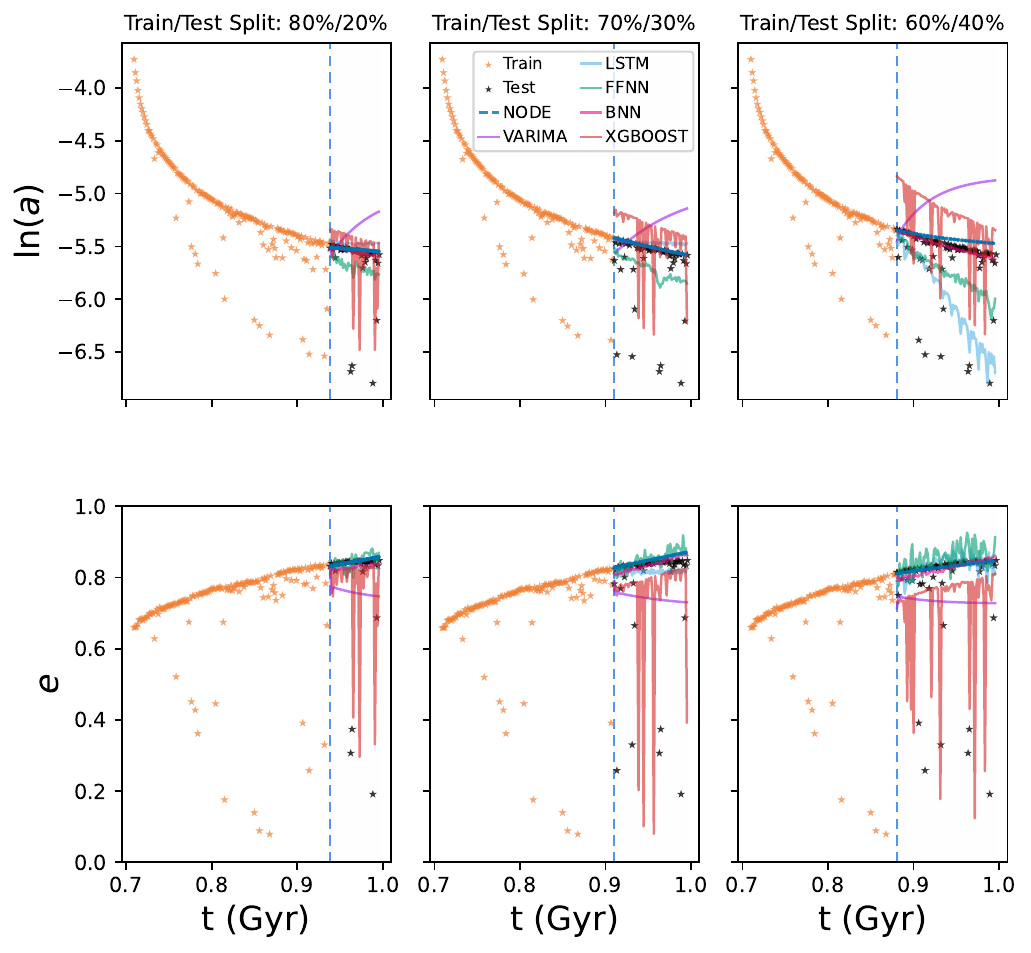}
       \caption{Forecasts of the Keplerian orbital elements $(\ln(a), e)$ for a single $N$-body simulation under several classical and neural models (FFNN, BNN, LSTM, XGBoost, VARIMA, NODE) at different chronological train/test splits. The blue dashed line indicates the start of the forecasting regime, and the semi-major axis $a$ is expressed in kpc (we plot $\ln(a)$). 
        Some models (e.g. XGBoost, VARIMA) exhibit visibly noisier trajectories due to the absence of inherent temporal smoothing; these are shown for completeness and comparison, and across all methods the forecast errors over this approximately linear interval are broadly comparable, with no clear performance winner and no systematic advantage for more flexible models such as the NODE or less flexible but strongly regularised models such as the BNN.}
    
    \label{fig:forecasting_models}
\end{figure*}

%%%%%%%%%%%%%%%%%%%%%%%%%%%%%%%%%%%%%%%%%%%%%%%%%%

% Don't change these lines
\bsp	% typesetting comment
\label{lastpage}
\end{document}